\documentclass{ws-p8-50x6-00}
\input epsf
\begin{document}

\title{Excitation of Nucleon Resonances}

\author{Volker D. Burkert}

\address{Thomas Jefferson National Accelerator Facility,\\ 
12000 Jefferson Avenue, Newport News, VA 23606, USA}

\maketitle

\abstracts{I discuss developments in the area of nucleon resonance excitation,
 aimed at putting our understanding of nucleon structure
 in the regime of strong QCD on a qualitatively new level. They involve 
 the collection of high quality data in various channels, 
 a more rigorous approach in the search for ``missing'' resonances, 
 an effort to compute some critical quantities in nucleon resonance 
 excitations from first principles, i.e. QCD , and a proposal focussed 
 at obtaining an understanding of a fundamental quantity in nucleon structure.}

\section{Introduction}

It is not easy to give an ``OUTLOOK'' talk after we have heard so many interesting
 new results, and most speakers talked already about plans for the future. So, I
 will be just speaking about a few selected aspects of nucleon resonance physics 
where progress may be possible in the next few years. I will
 point out a few areas where there have been significant advances recently,
 and where we may expect important progress soon. Finally, I want to present 
to the community a proposal for a focussed effort to study the $Q^2$ evolution 
of the generalized Bjorken integral from small to large distances. If
 successful, this would be an {\sl important milestone of nucleon structure studies}.

Electromagnetic production of mesons, which is what we are doing when studying 
resonances, may be crudely characterized by 3 regions representing different 
distance scales. At large distances, say 1 fm, we study nucleon
 properties near the surface. Nucleons and pions are the relevant 
degrees-of-freedom. Chiral perturbation theory describes many phenomena, and is linked 
to QCD via chiral symmetry and chiral symmetry breaking. At the other end, at very 
small distances, we probe the parton structure of the nucleon. 
Elementary quark and gluon fields are the relevant degrees of freedom used in 
perturbative QCD. This workshop has been mostly about the regime of 
intermediate
 distances, where the excitation of baryon resonances is prominent. Quarks and
 gluons are relevant; however, they interact more like constituent quarks and 
glue. This domain is currently addressed theoretically by quark 
models, flux tube models, QCD sum rules, instanton models, etc., with some 
success. The relationship of these approaches to QCD is often not fully 
developed, making it difficult to assess the accuracy of the model prediction.

The regions of different distance scales are likely not strictly separated 
from each other. This should provide areas of overlap where different theoretical 
approaches can be used to compute the same observables, thus allowing important checks 
of the range of validity of a specific approach. What I see as an important 
task for the community is to anchor more firmly these descriptions to the 
fundamentals of QCD, and finally come to an understanding of resonance 
phenomena and nucleon structure from the largest to the smallest distances 
within fundamental theory.

Nucleon structure studies are often associated with deep inelastic scattering where the
 interpretation of data in terms of the underlying degrees-of-freedom is 
usually more straightforward. However, quark structure function measurements 
and the test of asymptotic sum rules are only one small area of the nucleon 
structure to be explored, and certainly not the one most strongly 
related to QCD. Strong interaction plays no role in asymptotic sum rules, 
and the determination of quark structure
 functions is related to QCD only via secondary effects such as gluonic 
corrections. It is understanding the measured parton distributions that is 
the real challenge for QCD. 

Information on nucleon structure from formfactors or nucleon resonance 
excitations is much richer, albeit more difficult to interpret. 
Nevertheless these are quantities closer to the ``real'' world, and they 
need to be described and understood in terms of the underlying 
degrees-of-freedom if we want to make progress.

\section{The $\gamma N\Delta(1232)$ transition - from precision experiments to precision
calculations.}    

This is the region where we aim for precision. The $\Delta(1232)$ is the 
only resonance that is well separated from all the higher mass states. At 
low $Q^2$ the $\Delta(1232)$ has the largest cross section. 
There has been considerable 
progress in experiments and analyses over the past 5 years or so. The uncertainties 
in ratios of multipoles $R_{EM}= E_{1+}/M_{1+}$ and $R_{SM}=S{1+}/M_{1+}$ have been 
reduced by an order of magnitude since the early studies. 
This now allows sensible comparisons with model 
predictions.  One of the most noteworthy results is the quantitative realization that 
a description of the $N\Delta(1232)$ transition requires the inclusion of pions as 
effective degrees of freedom. A simultaneous description of both ratios is achieved 
only by models that include pion d.o.f\cite{lcsmith}. Also the total absorption cross sections for 
the $N\Delta(1232)$ transition is well described by models that include pion cloud effects.

A more precise description of quantities such as $R_{EM}$ in lattice QCD (LQCD) is long overdue. 
The only LQCD ``prediction'' of $R_{EM}$ is nearly a decade old\cite{leinweber} and ``predicts'' a 
value of $3 \pm 8\%$ at the photon point, while the experimental value is 
$(-2.75 \pm 0.50)\%$, where the experimental error is estimated generously. An order 
of magnitude smaller LQCD error is needed to have any impact here. A simple 
extrapolation of the computer performance using Moore's law, gives precisely 
the factor ten needed for progress. The next step would be to evolve this quantity 
in $Q^2$, and compute the magnitude of the magnetic transition multipole vs $Q^2$. 
This would mark real progress on the lattice!        

We also would like to know whether the apparent trend in the $R_{EM}$ data really 
indicates\cite{tiator} that there will be a zero crossing at $Q^2 \approx 4$GeV$^2$. This may give 
us a clue where leading order pQCD contributions may have some relevance. Since the 
signal/background at high $Q^2$ will be a lot smaller we also need to refine our analysis 
techniques and collect more data that give us more direct information on background 
contributions. Beam spin asymmetries as well as other polarization observables are 
needed to reduce the model-dependenc of the analysis techniques at high $Q^2$.

\section{The 2nd resonance region} 

The so called 2nd resonance region, comprising the mass range from 1.4 to 1.6 GeV, 
is of particular interest for nucleon structure studies. It contains 3 states,
 the $N(1440)P_{11}$ ``Roper'', the $N^*(1535)S_{11}$, and
 $N^*(1520)D_{13}$ states, all of which are highly interesting for the
 study of nucleon structure properties, and for the testing of basic symmetry
 properties.

\subsection{Mysteries of the Roper resonance $N(1440)P_{11}$}   

A natural candidate for detailed studies beyond the $\Delta(1232)$ would
 be the Roper resonance $N(1440)P_{11}$. However, more than 35 years after
 its discovery its structure is basically still unknown. The non-relativistic
 constituent quark model (nrCQM) puts its mass above 1600MeV, the photocoupling 
amplitudes are not described well, and the transition formfactors, 
although poorly determined, are far off. Relativized variations 
of the nrCQM improved the situation only modestly. To obtain a better 
description of the data a number of alternative models have been proposed. 
Does the Roper have a large gluonic component\cite{libuli}? Is it a small 
quark core with a large pion cloud\cite{cano}? Is it a nucleon-sigma molecule\cite{krewald}? 
Or, is it not a single resonance but two appearing in different 
reactions differently\cite{morsch}? These questions will be discussed at future 
workshops, however, it is crucial to get more precise electroproduction data, 
as it is the $Q^2$ dependence where the models differ strongly. From the model 
builders we must require that their models make predictions for the electromagnetic 
couplings and formfactors. 

There is also some good news: lattice QCD calculations are beginning to produce 
results for the mass of the Roper which may soon be accurate enough to have a 
real impact.   

\subsection{The $N^*(1535)S_{11}$ and $N^*(1520)D_{13}$, and the $[70,1^-]$ supermultiplet}

There is some good news from the constituent quark model. The slow fall-off of the 
transverse $N^*(1535)S_{11}$ transition formfactor, which has been a problem for
 model builders for a long time, is now quite well described by the CQM using a potential 
containing a Coulomb form and a linear term\cite{giannini}. At the same time the 
$A_{1/2}$ amplitude of the $D_{13}$ is described as well, while there remains a 
large discrepancy for the $A_{3/2}$ amplitude at small $Q^2$. Could this 
be explained by pionic contributions which then would have to contribute to the 
helicity nonconserving (nonleading) term but not to the helicity conserving 
(leading) amplitude? Calculations that include pionic contributions explicitly 
are needed to answer this question. 

Another piece of good news comes from LQCD. As already discussed at the  previous 
workshop\cite{sasaki}, mass predictions for the lowest $N^*$ state with 
negative parity agree well with the experimental values. The obvious next 
step would be to compute the $A_{1/2}$ amplitude for that state at the photon 
point in LQCD. 

For a better understanding of the Roper as well as the $N^*(1520)D_{13}$, data 
in the n$\pi^+$ channel are crucial to obtain more complete isospin information. 
Also, beam spin asymmetry measurements will give information about the background 
amplitudes which are especially important in that mass region. Such data have been 
taken and are currently being analyzed\cite{kjoo}.

The ordering of excited states according to the $SU(6)\otimes O(3)$ symmetry 
group and the assumption that excitations are due to a single quark transition 
(SQT) allows predictions for a large number of states belonging to the same 
supermultiplet based on only three known amplitudes. In the case of the 
$[70,1^-]$ the $N^*(1535)S_{11}$ and the $N^*(1520)D_{13}$ may serve that 
purpose. These are the only states in this multiplet whose transition amplitudes 
have been measured with some accuracy. This allows tests of the SQT assumption, 
and how the symmetry will break down as a function of the distance scale. While 
the predicted photocoupling amplitudes are in quite good agreement with the data, 
there are not enough data at finite $Q^2$ to test this simple model at shorter 
distances. The lack of data for two of the prominent states, the 
$\Delta(1620)S_{31}$ and $\Delta(1700)D_{33}$, is largely due to the complete 
lack of data in the $N\pi\pi$ channel. Also, amplitudes for neutron resonances 
are absent for all states. This situation will hopefully change soon with new 
data from CLAS\cite{ripani}.

\section{Missing baryon resonances}

Understanding the fundamental structure of baryons remains the main focus 
of the $N^*$ progam. There is now a significant effort underway 
to search for some of the states predicted by the symmetric quark model\cite{isgkar} 
that have not been seen 
in $\pi N$ scattering. The importance of this effort lies in the fact that 
these states can tell us much about internal baryon structure. For example, 
models that do not have approximate SU(6) symmetry may not predict some or 
even many of these states to exist\cite{kirchbach}. 
Some of these states are predicted to couple 
to $\Delta\pi$, $N\omega$, $Y^*K$, and other hadronic channels, as well as to 
photons\cite{caprob}. Photo- or electroproduction may therefore be the only way to search 
for some of these states. Experiments at GRAAL, JLab, ELSA (Crystal Barrel), 
Spring-8, and BEPC have begun a vigorous search employing large acceptance 
detectors\cite{missres}. This effort is accompanied by a theoretical effort to understand 
how these resonances might show up in experimental observables\cite{bennhold,oh}. 

There are indications for one or even two of such states which have been 
discussed at this workshop. Some of this evidence is, however, due to 
improvements  that model curves show in comparison to data in case such states 
are included. Clearly, this is not sufficient. Other partial wave contributions 
need to be tested and excluded. For the evidence to be fully convincing, partial 
wave analyses must be done that seek to analyse such states in the energy-dependence 
of partial-wave amplitudes and their phase motion. 

The strangeness sector offers excellent prospects in the search for missing
states. Hyperon resonances are more narrow than states made of u and d quarks
only, and they can be separated more easily from other overlapping 
resonances\cite{nefkens}.

Another kind of ``missing'' baryons are the gluonic excitations or ``hybrid'' 
states where the ``glue'' or flux tubes are excited and produce a $|q^3G>$ state. 
They have been predicted in bag models and flux tube models. Lattice QCD predicts 
such states in the meson sector. They are likely expected in the baryon sector as 
well, although no LQCD calculations have been performed. 
In distinction to the meson sector no exotic quantum numbers are expected in the 
baryon sector. This will make it experimentally more difficult to identify 
gluonic excitations. QCD sum rules\cite{kissl}, flux tube\cite{page}, and bag 
models\cite{haqq} predict the 
lowest gluonic states to be $P_{11}$ or $P_{13}$ with masses between 1.5 GeV 
for bag models and QCD sum rules, and 1.8 - 1.9 GeV for the flux tube model\cite{page}. 
Possible signatures
could be the overabundance of states, unusual decay channels, form 
factors which are different from the 3-quark sector due to the larger sizes of $|q^3G>$ 
states, and different threshold behavior due to different 
$SU(6)\otimes O(3)$ assignment.  Another possibility is the production 
of hybrid states in the gluon rich environment of $J/\psi$ decays\cite{zou}.

None of these signatures alone will be convincing. It will take various pieces of 
evidence, and a good understanding of these states within models 
that treat 3-quark states and gluonic states on an equal footing, to have 
sufficient confidence in any discovery in this area.


\section{The nucleon spin integral from small to large distances - A proposal for the next 5 years} 

Coming back to the goals outlined in the introduction one may ask what quantities 
are most directly 
accessible to a description within fundamental theory. As ``fundamental'' I 
would characterize exact sum rules, such as the GDH and Bjorken sum rules, QCD, 
pQCD, and chiral perturbation theory. I will argue that 
$\Delta\Gamma^{pn}_1(Q^2)= \int{[g^p_1(x,Q^2)-g^n_1(x,Q^2)]dx}$ is such a 
quantity.   

What would be the significance of such a project? Why is it important, and why
 should the $N^*$ community be involved in this?  Clearly, from a physics 
perspective such a project, if successful, would be a milestone, as it would 
{\sl mark the first time a 
fundamental quantity of nucleon structure is described by fundamental theory 
from small to large distances}, a worthwhile goal of nucleon structure physics, 
and worth a serious effort by the community. First, the expertise of the 
$N^*$ community is important as nucleon resonances make significant
 contributions to the spin integral at medium and large distances\cite{buli,buriof}. Second, 
such a project provides a focus for the community to solve a 
fundamental problem. Third, the description of the resonance contributions to 
the first moment in LQCD may be the biggest effort, and there are proposals from within 
the community to 
have significant computing resources available for nucleon structure studies 
in the next five years, that can be brought to bear on such a project.

\begin{figure}
\vspace{65mm} 
\centering{\includegraphics{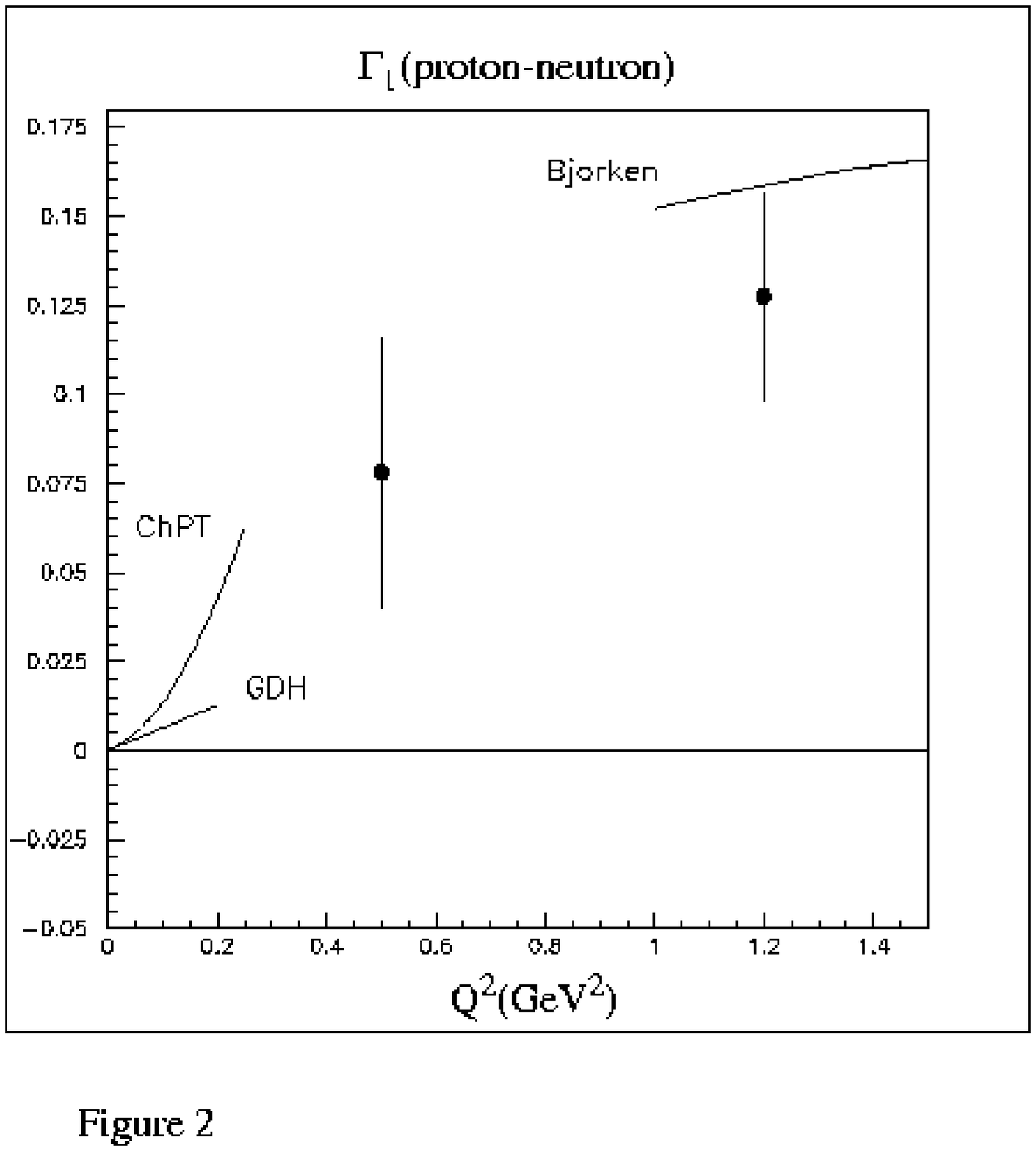}}
\centering{\includegraphics{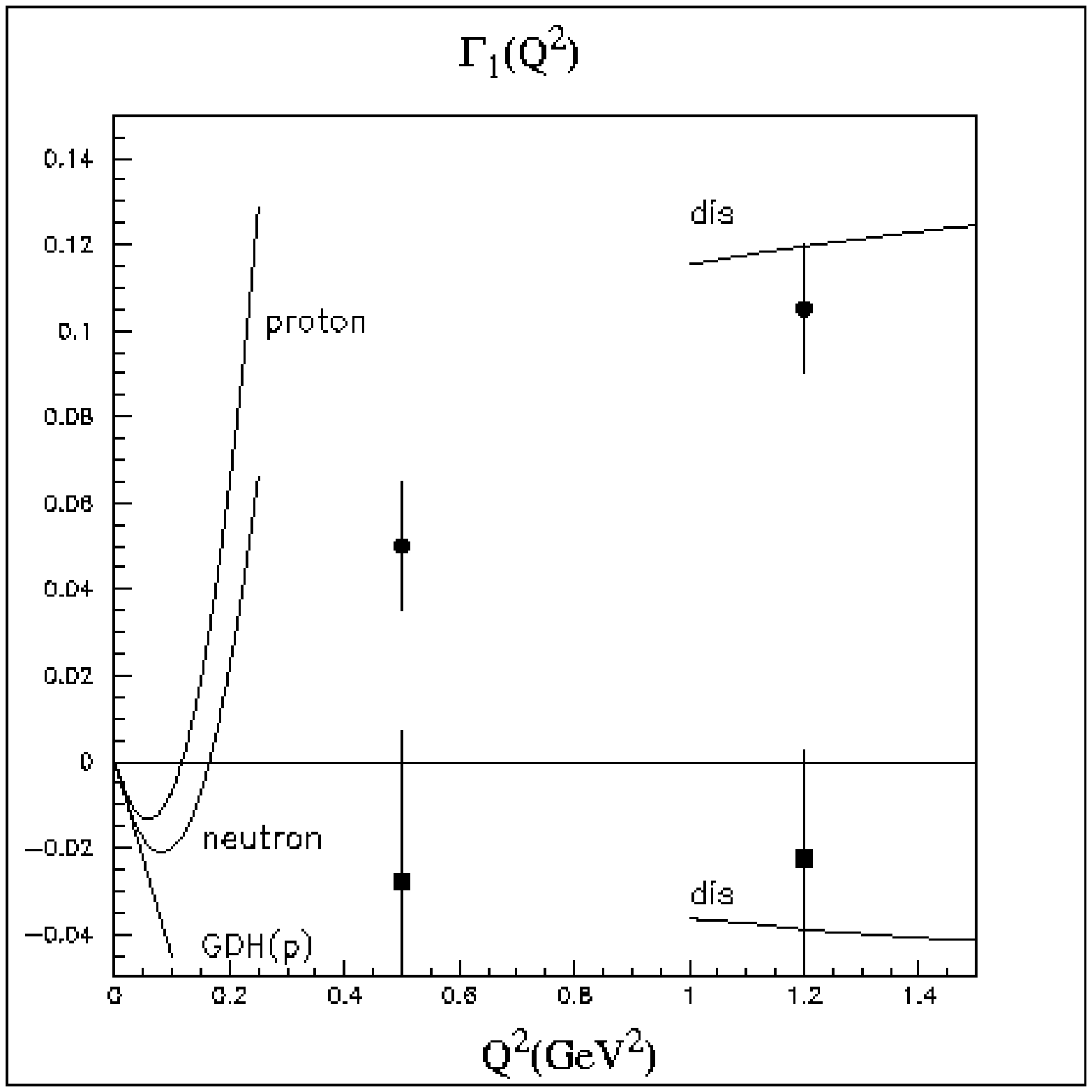}}
\caption{\small First moment of the spin structure 
function $g_1(x,Q^2)$ for the 
proton and neutron (left), and for the proton-neutron 
difference (right). The curves above $Q^2=1$GeV$^2$ are pQCD evolutions of 
the measured $\Gamma_1$ for proton and neutron, and the pQCD evolution for 
the Bjorken sum rule, respectively. The straight lines near $Q^2=0$ are the 
slope given by the GDH sum rule. 
The curves at small $Q^2$ represent the NLO HBChPT results.}.
\label{fig:gamma1}
\end{figure}

\subsection{What is the experimental and theoretical situation?} 

The experiments to measure polarized structure function $g_1(x,Q^2)$ in a large 
$Q^2$ range are far along as has been reported at this 
conference\cite{minehart,chen}. The deep inelastic regime has been 
studied for decades, and good data are available for $Q^2 > 1.5 GeV^2$ mostly for 
the proton but also for the neutron. Experiments at JLab in CLAS and in Hall A are near
 to final results for the range in $Q^2 = 0.1 - 1.0$ GeV$^2$. These data currently
 require an extrapolation at small x which adds a small systematic error in the low 
$Q^2$ range, however, a significant uncertainty at $Q^2 > 1$ GeV$^2$. 
This situation is changing with the new data taken with CLAS in the energy range 
from 1.6 - 5.75 GeV, and in Hall A with an upcoming experiment at very small 
$Q^2$. Also, uncertainties in the extraction of the neutron contribution   from 
measurements on $^3He$ require an improved treatment of the nuclear effects at 
small $Q^2$ where uncertainties are significant.

Within this year the first complete information on $\Gamma^p_1(Q^2)-\Gamma_1^n(Q^2)$ 
should be available in a $Q^2$ range from 0.1 - 1 GeV$^2$ from JLab experiments. At 
the same time, information on the proton and the neutron separately will be available as
 well.  
The current theoretical situation is illustrated in Figure {\ref{fig:gamma1}}. 
The left hand panel is 
for the proton and for the neutron separately. The high $Q^2$ behavior has been
 measured, and is known to approach a constant value. The asymptotic 
behavior has been evolved to lower $Q^2$ in perturbative QCD to order $\alpha_s^3$.
 This is shown by the lines labelled ``dis''. At the low $Q^2$ end we have 
the GDH sum rule believed to be valid at the photon point. It also defines the slope
 of $\Gamma_1(Q^2 \rightarrow 0)$. The slope is negative for both proton and neutron.
 Heavy Baryon Chiral Perturbation Theory (HBChPT) has been used\cite{ji} to evolve the GDH 
integral to finite $Q^2$. The curves are from NLO calculations.
 Unfortunately, for the proton and neutron this expansion appears to break down 
already at very small $Q^2$. A potential problem in these calculations is the treatment
 of the $\Delta(1232)$. To avoid this problem we take the 
proton-neutron difference, where this contribution is not present. The 
result is a dramatic improvement in the low $Q^2$ description of the apparent trend 
of the data\cite{burk}. $Q^2$ values up to 0.25 GeV$^2$ or higher might be 
reachable in this quantity. Taking the proton-neutron difference 
$\Delta\Gamma^{pn}_1$
is also suggested from the  behavior in the deep inelastic regime where the Bjorken
sum rule\cite{bjorken} establishes an important constraint for the absolute 
normalization of the first moment, which has been verified experimentally within 
5-10\% 
 
The combination of two fundamental sum rules at the opposite sides of the distance 
scale, with the pQCD evolution at small distances and the ChPT evolution at large 
distances provide powerful constraints for the $Q^2$ evolution of that quantity 
throughout the entire distance scale.  
This provides a unique opportunity to describe $\Delta\Gamma^{pn}_1$ within 
fundamental theory. This may require going to higher order in ChPT, and to lower 
$Q^2$ in the Operator Product Expansion of pQCD. In addition, it may be nessecary to employ 
lattice QCD to cover the intermediate distance scale and provide an overlap with
 both the higher and the lower $Q^2$ domains. 
These calculations must be confronted with precise measurement of resonance 
contributions to the 
spin integral. 

\vspace{-0.2truecm}
\section*{Closing remarks}

As the last speaker of this workshop I have the honor, and the pleasant 
obligation and opportunity, to express the gratitude of the participants 
at this workshop 
to the organizing committee and its chairman, Dieter Drechsel, for an 
excellent scientific program, organized in a most friendly 
atmosphere, for the superb food, and for providing the 
opportunity for in-depths discussions with collegues and friends. For this 
we say

\vskip0.3cm
 
   \centerline{\Large D a n k e ~ ! }

\section*{In memoriam}

One of our best, who is no longer with us, Nimai Mukhopadhyay, a friend to many 
of us, and a champion of nucleon structure studies and of baryon resonances, 
was sorely missed at this workshop. Nimai made many important contributions to this 
field, and organized workshops like this one. We can honor his name by making baryon 
resonances an even more visible part of nucleon structure studies in the years to come.


\end{document}